\documentclass[aps,prl,twocolumn,showpacs]{revtex4-1}
\usepackage{epsfig}
\usepackage{graphicx}
\usepackage{bm}
\usepackage{amssymb}
\usepackage{amsmath}
\usepackage{amsthm}
\usepackage{color}
\usepackage{ulem}

\begin{document}
\title{Electric field driven magnetic domain wall motion in ferromagnetic-ferroelectric
heterostructures}
\author{Ben Van de Wiele$^1$, Lasse Laurson$^2$, K\'{e}vin J. A. Franke$^3$, and Sebastiaan van
Dijken$^3$}
\affiliation{$^1$Department of Electrical Energy, Systems and Automation,
Ghent University, Ghent B-9000, Belgium}
\affiliation{$^2$COMP Centre of Excellence, Department of Applied Physics, 
Aalto University, P.O. Box 11100, FIN-00076 Aalto, Espoo, Finland}
\affiliation{$^3$NanoSpin, Department of Applied Physics,
Aalto University School of Science, P.O. Box 15100, FI-00076 Aalto, Finland}

\begin{abstract}
We investigate magnetic domain wall (MDW) dynamics induced by applied electric fields in
ferromagnetic-ferroelectric thin-film heterostructures. In contrast to conventional driving
mechanisms where MDW motion is induced directly by magnetic fields or electric currents, MDW motion
arises here as a result of strong pinning of MDWs onto ferroelectric domain walls (FDWs) via local
strain coupling. By performing extensive micromagnetic simulations, we find several dynamical
regimes, including instabilities such as spin wave emission and complex transformations of the MDW
structure. In all cases, the time-averaged MDW velocity equals that of the FDW, indicating the
absence of Walker breakdown.  
\end{abstract}
\maketitle

Magnetic domain wall (MDW) dynamics in nanoscale ferromagnetic wires and strips, as well as in thin
films, is a subject of  major technological importance for the operation of potential future
magnetic memory \cite{PAR-08} and logic devices \cite{ALL-05,JAW-09}. 
While current efforts to construct such devices mostly focus on spin-polarized electric currents
\cite{BER-96, SLO-96} or applied magnetic fields \cite{SHI-11} as the driving force, a promising
low-power alternative has been demonstrated in recent experiments \cite{LAH-12} where {\it
electric} fields move the MDWs in ferromagnetic-ferroelectric heterostructures.  In such
configurations, the MDWs in the ferromagnetic layer are strongly pinned onto ferroelectric domain
walls (FDWs) in the ferroelectric sublayer via elastic interactions. Consequently, when an applied
electric field displaces the FDWs, the MDWs are dragged along.

From a fundamental physics point of view, the question of the nature of this driving protocol is
very important. Indeed, the electric field driving mechanism of MDWs differs substantially
from the more conventional driving modes where either a magnetic field or spin-polarized electric
current are used to move MDWs. While the effect of an applied electric field on magnetic field
\cite{SCH-12, CHI-12, BAU-13} and spin-polarized current \cite{LEI-13} driven MDW motion has been
considered, the nature of fully electric field driven MDW dynamics in ferromagnetic-ferroelectric
heterostructures remains elusive. 

In this Letter, we present a detailed numerical study of electric field driven MDW dynamics, including the short time scale details which have so far not been
accessible experimentally. We consider two different 90$^\circ$ MDWs, one being magnetostatically
{\it charged} and the other {\it uncharged}. Our results highlight the different nature of the
electric field driving mechanism as compared to well-known magnetic field and
electric current driven MDW motion: For all applied FDW velocities, the MDW moves along with a
time-averaged velocity equal to the FDW velocity. Thus, the sharp decrease of the average velocity
associated with Walker breakdown \cite{SHI-11, SCH-74, NAK-03, BEA-05, HAY-07} in magnetic field and
electric current driven MDW dynamics is absent. For small FDW velocities, the MDW coupled to it is
found to follow the moving FDW nearly quasistatically, without significant changes in the internal
MDW structure. Above a threshold velocity,  
this close-to-quasistatic behavior breaks down, with various instabilities occurring depending on
the MDW type (charged or uncharged) and the material parameters. For uncharged MDWs these
instabilities comprise oscillatory MDW motions or complicated transformations of the MDW structure,
while the emission of regular spin waves is observed for charged MDWs.

\begin{figure*}[t!]
\includegraphics[width=\textwidth]
{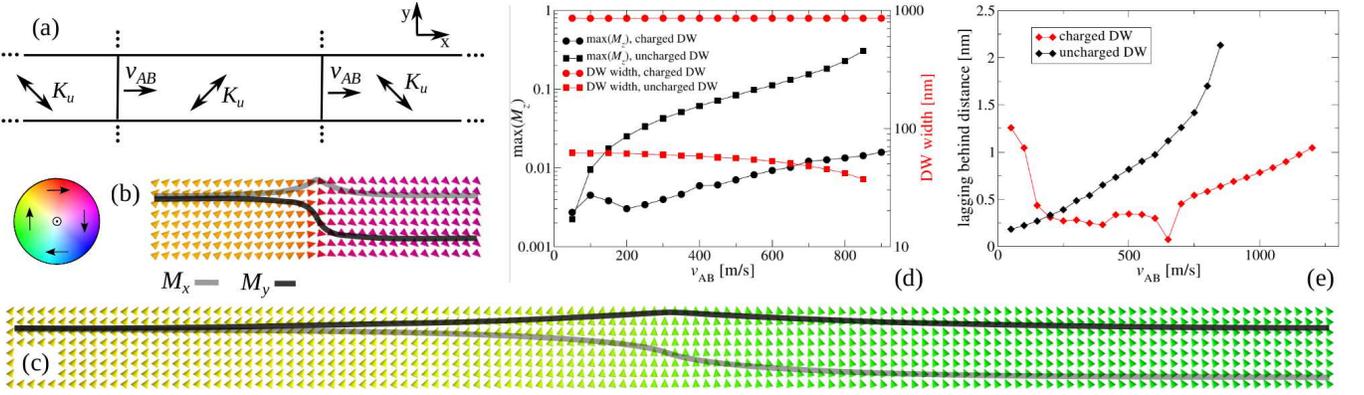}
\caption{(color online) (a) Sketch of the micromagnetic simulation geometry. The system is
discretized in two dimensions using finite difference cells of
3.125$\times$3.125$\times$15\,nm$^3$. The micromagnetic structures of the equilibrium uncharged and
charged 90$^{\circ}$ MDWs are shown in (b) and (c), respectively, along with the color code used. 
Each arrow represents the locally averaged magnetization over a 12.5$\times$12.5$\times$15\,nm$^3$
volume, i.e. the scale 
of the plots in (b) and (c) is identical. (d) Evolution of the maximum out-of-plane magnetization
component and MDW width with  AB velocity. (e) Lagging behind distance in the quasistatic regime
with $v_{AB} < v_{th}$ \cite{comment1}. For the reference material parameters used here, 
$v_{th} \approx 850$ m/s for the uncharged MDW and $v_{th} \approx 1200$ m/s for the charged MDW.}
\label{fig:fig1}
\end{figure*}

In ferromagnetic-ferroelectric thin-film heterostructures, the anisotropy of the ferromagnetic layer
is laterally modulated via local strain transfer from ferroelectric domains and inverse
magnetostriction. If polarization rotation between ferroelectric domains is less than 180$^\circ$
(i.e. if the domain pattern consists of FDWs that are both ferroelectric $and$ ferroelastic),
minimization of the anisotropy energy can lead to full imprinting of ferroelectric domains into the
ferromagnetic layer \cite{LAH-11-1, LAH-11-2, CHO-12, LAH-12-2}. Importantly, abrupt rotation of the
ferroelectric polarization at FDWs and the concurrent instant change of magnetic anisotropy in the
adjacent ferromagnet strongly pins the MDWs onto their ferroelectric counterparts. The reverse
effect, i.e. modulation of the ferroelectric properties due to magnetization reversal in the
ferromagnetic film, does not occur due to a pronounced asymmetry in the strain coupling mechanism.
The maximum strain that can be transferred from a ferroelectric domain is given by the elongation
of the structural unit cell. For archetypical tetragonal BaTiO$_3$ at room temperature, this is
1.1\%. Strain transfer from a ferromagnetic material to a ferroelectric layer via magnetostriction,
on the other hand, is typically several orders of magnitude smaller \cite{HAN-00}. As a result,
ferromagnetic effects on the dynamics of the ferroelectric sub-system can be safely neglected, and
electric field induced MDW propagation is accurately modeled using a moving magnetic anisotropy
boundary (AB) in the ferromagnetic layer. 

In our micromagnetic simulations, an AB separating two regions of uniform uniaxial anisotropies is
moved with constant velocity $v_{AB}$ (Fig. \ref{fig:fig1}(a)), corresponding to the propagation
velocity of the underlying FDW. The angle between the magnetic anisotropy axes is 90$^\circ$, which
in practice can be obtained by strain coupling to in-plane domains of BaTiO$_3$ \cite{LAH-11-1, LAH-11-2}. In an experiment, $v_{AB}$ would be controlled by the magnitude of an applied electric
field \cite{LIT-55, JO-09, SON-13}. Coarse-grained Monte Carlo simulations indicate that
ferroelectric DW speeds of up to several km/s are possible \cite{SHI-07}. Our simulations are
performed with the GPU-based micromagnetic simulator MuMax \cite{VAN-11}. To study the time
evolution of the magnetization ${\bf M}({\bf r},t)$, we solve the Landau-Lifshitz (LL) equation, 
\begin{eqnarray}
\label{eq:1}
\frac{\partial {\bf M}}{\partial t} & = & -\frac{\gamma}{1+\alpha^2} 
{\bf{M}}\times {\bf H}_{eff} \\ \nonumber
& & -\frac{\alpha \gamma}{M_s(1+\alpha^2)} 
{\bf M}\times({\bf M}\times{\bf H}_{eff}), 
\end{eqnarray}
where ${\bf H}_{eff}$ is the effective magnetic field (with contributions from the exchange,
anisotropy and demagnetizing fields), and $\gamma$ is the gyromagnetic ratio. The reference sample
in this study is a 15\,nm thick Co$_{60}$Fe$_{40}$ layer on top of a BaTiO$_3$ substrate,
corresponding to the experimental system in \cite{LAH-11-1}, with saturation magnetization
$M_s$=1.7$\times10^6$\,A/m, uniaxial magnetic anisotropy strength $K_{u}$=1.7$\times10^4$\,J/m$^3$,
exchange constant $K_{ex}$=2.1$\times10^{-11}$\,J/m, and damping constant $\alpha$=0.015. All
material parameters are varied around these values to investigate their effect on the observed MDW
dynamics.  By considering periodic boundary conditions in both $x$- and $y$-direction, we mimic MDW 
propagation in an infinite film.  The simulation window is restricted to two 6.4\,$\mu m$ 
wide domains with orthogonal anisotropy axes over a length of 200\,nm, see Fig. \ref{fig:fig1}(a). 
The resulting 90$^\circ$ MDWs can be of two different types, magnetostatically charged or uncharged, 
depending on the magnetization directions in the domains \cite{FRA-12}. The widths of these two MDW 
types differ substantially as illustrated by their micromagnetic structures in Figs. \ref{fig:fig1}(b) 
and (c). We consider both cases separately and show that their dynamic behavior is very different.  

\begin{figure*}[t!]
\includegraphics[width=\textwidth,clip]{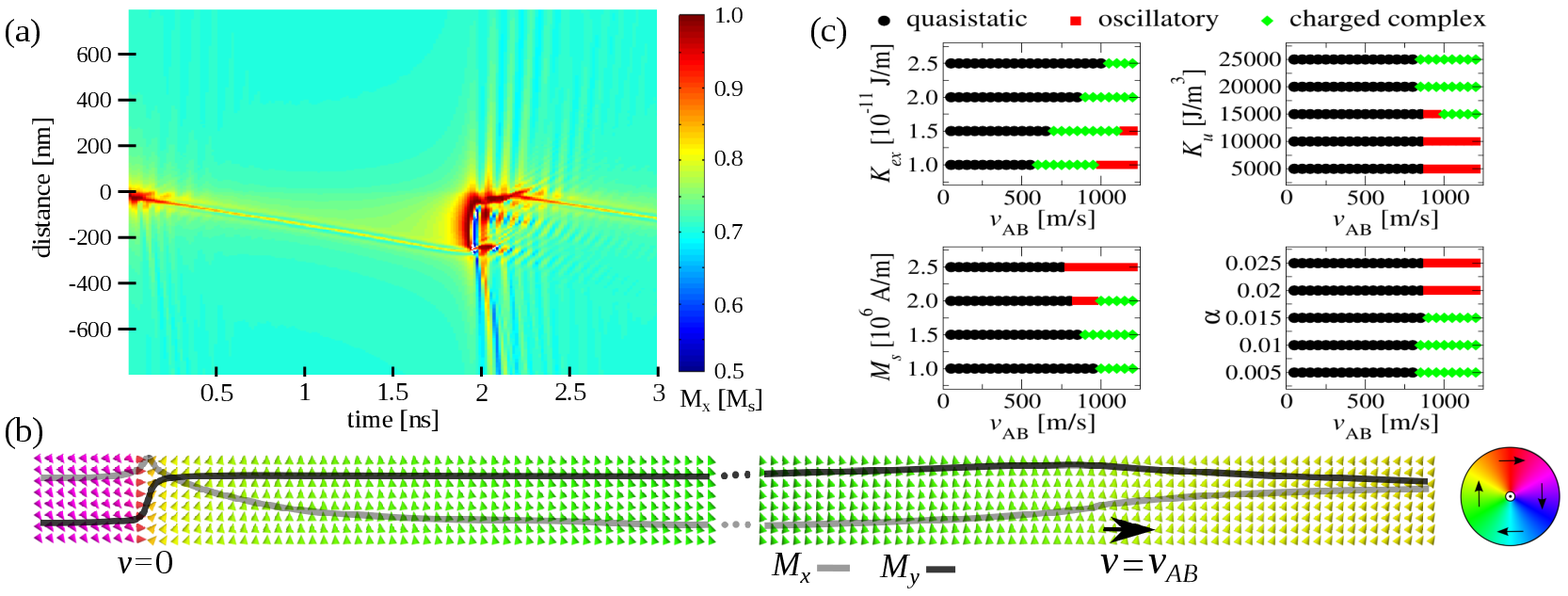}
\caption{(color online) (a) Local $x$-component of the magnetization of an uncharged MDW displaying 
``oscillatory'' behavior above $v_{th}$. The position of the domain wall is shown in the moving 
framework of the AB, with $v_{AB} = 1000$ m/s. The uncharged MDW is initially moving at a reduced velocity of about 900 m/s for the first 2 ns, after which it is abruptly pulled back to the AB. (b) Magnetization structure of an uncharged MDW 
that transformed into a ``charged complex'' MDW ($v=0$) and a charged MDW ($v=v_{AB}$). (c) 
Phase diagrams showing the effects of the micromagnetic parameters on $v_{th}$ and the type 
of MDW dynamics (``quasistatic'', ``oscillatory'', or ``charged complex'', see text for details). 
From top-left to bottom-right the parameters that are varied with respect to their reference 
values are the exchange constant, uniaxial magnetic anisotropy strength, damping constant, and 
the saturation magnetization.}
\label{fig:fig2}
\end{figure*}

For small imposed $v_{AB}$, both the charged and uncharged MDW follow the motion of the AB. While
the magnetization of the uncharged DW is completely in-plane at rest, an out-of-plane magnetization
component develops and the DW width reduces with increasing $v_{AB}$ (Fig. \ref{fig:fig1}(d)). These
deformations are similar to those observed during magnetic field and current driven MDW motion in
magnetic nanowires and strips, where the appearance of an out-of-plane magnetization component and a
narrowing of the MDW are precursors of Walker breakdown \cite{SHI-11, SCH-74, NAK-03, BEA-05,
HAY-07}. In addition, the uncharged MDW lags slightly behind the AB by a distance that increases
with $v_{AB}$ (Fig. \ref{fig:fig1}(e)). For higher AB speeds, a breakdown of quasistatic MDW motion
occurs at a threshold velocity $v_{th}$. In comparison, the internal structure of the charged MDW is
much more robust against dynamic deformations, as illustrated by the negligible out-of-plane
magnetization and nearly constant MDW width in Fig. \ref{fig:fig1}(d). 

The dynamics of the uncharged MDW above the threshold velocity $v_{th}$ depends on the material
parameters.  Two possible scenarios are observed in our simulations. In the first, termed here as
``oscillatory", the MDW first lags increasingly behind the AB until a maximum distance is reached
(about 200 nm for the reference parameters). At that point, the MDW is abruptly pulled back to the
AB (a process that also involves the emission of spin waves), after which the MDW starts to lag
behind the AB again (Fig. \ref{fig:fig2}(a)). This cycle of events is repeated continuously. The
second scenario, labeled as ``charged complex", is significantly different. In this case, the
initially uncharged MDW transforms by creating two oppositely charged MDWs. After this, one of the
charged MDWs follows the AB, while a MDW complex comprising the original uncharged MDW and the other
charged MDW is left behind. The MDW complex subsequently slows down and eventually stops moving
since it is no longer pinned onto a moving FDW (the AB in our simulations). This behavior is somewhat
analogous to recent observations of a break-up of the compact domain wall structure in wide submicrometer wires \cite{ZIN-11}. An example of the resulting configuration is shown in Fig. \ref{fig:fig2}(b). This process obeys the principle of
charge conservation, i.e. the net charge of the newly created charged MDWs is zero. Although the
deformations of an uncharged MDW below $v_{th}$ are reminiscent of Walker breakdown, the physical
consequences above $v_{th}$ are different compared to magnetic field or current driven magnetic
systems. In the latter cases, an abrupt decrease in the time-averaged MDW speed is often observed
beyond breakdown due to magnetization precession within the MDW. The average speed of an initially
uncharged MDW that is strongly coupled to a FDW equals the FDW velocity, and it moves either in an
``oscillatory" fashion or as a charged MDW after the transformation.

\begin{figure}[t!]
\includegraphics[width=\columnwidth]{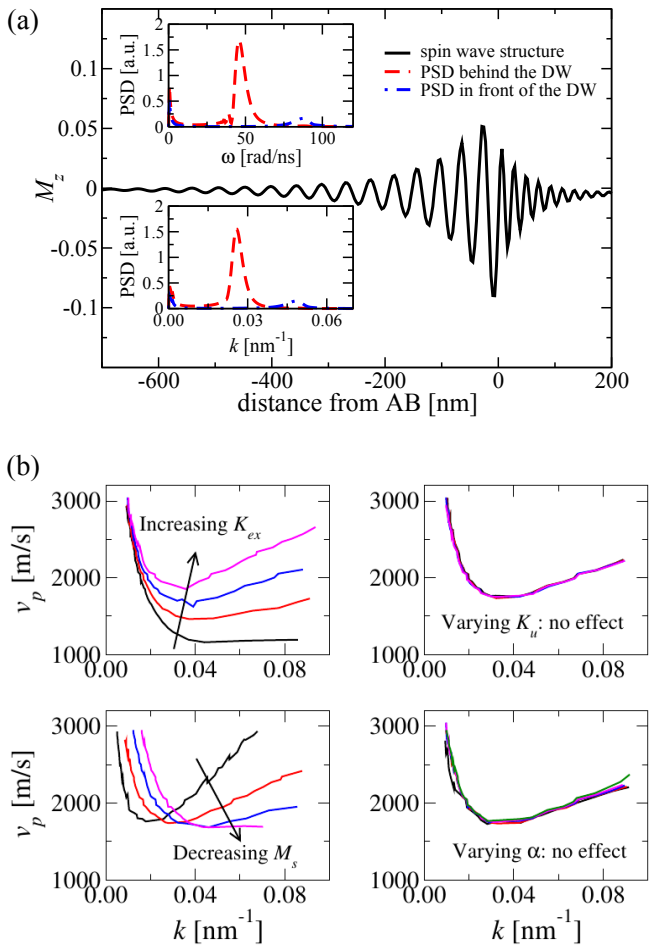}\\
\caption{(Color online) (a) Typical spin waves emitted by a moving charged MDW for $v_{AB} >
v_{p,min}$ along with the corresponding spatial and temporal power spectral densities (PSD) of the
waves. Here, the reference material parameters and $v_{AB}$ = 1800\,m/s are used. (b) Phase velocity
$v_{p}$ versus wave number $k$. The material properties of the reference sample are used, except for
the parameter that is varied using the same values as in Fig. \ref{fig:fig2}(c). The threshold
velocity for spin wave emission corresponds to the minimum of the curves. $v_p$ equals $v_{AB}$ and
determines the wave vector of the wave moving in front (largest $k$) and behind the MDW (smallest
$k$).}
\label{fig:fig3}
\end{figure}

The dependence of the threshold velocity $v_{th}$ and the type of dynamic behavior on the
micromagnetic parameters are illustrated by the phase diagrams of Fig. \ref{fig:fig2}(c). The
maximum distance between a moving uncharged MDW and the AB in the small imposed $v_{AB}$ regime (Fig. \ref{fig:fig1}(e)), i.e. the ``stiffness" of the pinning potential that is created by the AB, determines the dynamics
above $v_{th}$: Material parameters that give rise to a large lagging behind distance in the low
velocity regime lead to ``oscillatory" behavior above $v_{th}$, while small MDW-AB distances below
$v_{th}$ imply a ``charged complex" scenario for $v_{AB}>v_{th}$ (Fig \ref{fig:fig2}(c)). The
threshold velocity $v_{th}$ strongly depends on the exchange constant, with weaker exchange
interactions resulting in smaller $v_{th}$. Thus, in order to experimentally observe non-trivial
high velocity dynamic effects in ferromagnetic-ferroelectric heterostructures, it would be favorable
to use magnetic materials that combine a small exchange constant 
and large magnetostriction, with the latter needed in order to have a sufficiently strong pinning
of the MDWs onto their ferroelectric counterparts.

The dynamic behavior of charged MDWs is different from that of uncharged MDWs at high AB velocities.
 The much larger width of the charged MDW implies that it can move fast without encountering major
instabilities. This is due to the much lower rate of spin rotation needed to move the wide charged
MDW, as compared to the narrow uncharged MDW. The charged MDW starts to emit spin waves
when $v_{AB}$ exceeds the minimum spin wave phase velocity. A typical wave profile of a charged MDW
at $v_{AB}$ = 1800\,m/s is shown in Fig. \ref{fig:fig3}(a). The spin waves that are emitted have a
phase velocity $v_p$ which equals $v_{AB}$, but their well-defined wave number $k$, and thus their
frequency, is different in front and behind the moving MDW (insets of Fig. \ref{fig:fig3}(a)). This
behavior is similar to recent numerical observations of fast MDW dynamics in magnetic nanotubes
\cite{YAN-11}. 

The emitted spin waves in front and behind the MDW are characterized by dispersion graphs ($v_p$
versus $k$), which we obtained by considering different AB velocities.  The dependence on material
parameters (same values as in Fig. \ref{fig:fig2}(c)) is summarized in Fig. \ref{fig:fig3}(b). The
threshold velocity for spin wave emission corresponds to the minimum spin wave phase velocity
$v_{p,min}$. At this velocity, the spin waves moving in front and behind the MDW have the same wave
number and frequency. From Fig. \ref{fig:fig3}(b) it is clear that the exchange constant $K_{ex}$
has a large influence on $v_{p,min}$.  The saturation magnetization $M_s$ does barely influence
$v_{p,min}$, but it influences the corresponding wave number. The dispersion properties of the spin
waves emitted in front of the MDW are much more affected than those behind the MDW when $K_{ex}$ and
$M_s$ are varied. The strength of the uniaxial magnetic anisotropy $K_{u}$ and the magnetic damping
constant $\alpha$ have no effect on the emission spectra. The independence of $K_{u}$ is caused by the cancellation of two opposing effects, one related to the emission of spin waves (a stronger AB pumps more energy into the spin waves) and another due to the stiffness of the medium in which the spin waves propagate (larger anisotropy implies a stiffer medium). The negligible influence of $\alpha$ on the dispersion properties of spin waves is well known and often utilized in micromagnetic studies.

To summarize, we have studied electric field induced MDW motion in ferromagnetic-ferroelectric 
heterostructures. The driving force, which can be modeled as an anisotropy boundary moving in the
ferromagnetic layer, provides a mechanism for MDW dynamics exhibiting properties that are
fundamentally different from magnetic field and electric current driven MDW motion. Depending on the
MDW type and material parameters, spin wave emission and MDW transformations are found for high
driving velocities. Due to the robustness of this driving mechanism, manifested by the absence of
the Walker breakdown, electric field driven MDW motion could open up exciting opportunities in the
design of low power spintronics applications such as magnetic memory and logic devices.

{\bf Acknowledgments}. We thank Mikko Alava for a critical reading of the manuscript.
This work has been supported by the Flanders Research 
Foundation (B.V.d.W.), the Academy of Finland through a Postdoctoral Researcher's 
Project (L.L., project no. 139132), an Academy Research Fellowship (L.L., project 
no. 268302), an Academy project (S.v.D., project no. 260361) and via the Centres 
of Excellence Program (L.L., project no. 251748), as well as by the Theory Programme 
of Helsinki Insitute of Physics (L.L.), by the Finnish Doctoral 
Program in Computational Sciences (K.J.A.F.) and by the European Research 
Council (S.v.D., ERC-2012-StG 307502-E-CONTROL)


\begin{thebibliography}{10}
\bibitem{PAR-08}
S. S. P. Parkin, M. Hayashi, and L. Thomas,
Science {\bf 320}, 190 (2008).
\bibitem{ALL-05}
D. A. Allwood, G. Xiong, C. C. Faulkner, D. Atkinson, D. Petit, and R. P. Cowburn,
Science {\bf 309}, 1688 (2005).
\bibitem{JAW-09}
J. Jaworowicz, N. Vernier, J. Ferr\'e, A. Maziewski, D. Stanescu, D. Ravelosona, A. S. Jacqueline, C. Chappert, B. Rodmacq, and B Di\'eny, Nanotechnology {\bf 20}, 215401 (2009).
\bibitem{BER-96}
L. Berger,
Phys. Rev. B. {\bf 54}, 9353 (1996).
\bibitem{SLO-96}
J. C. Slonczewski,
J. Magn. Magn. Mater. {\bf 159}, L1 (1996).
\bibitem{SHI-11}
J. Shibata, G. Tatara, and H. Kohno,
J. Phys. D: Appl. Phys. {\bf 44}, 384004 (2011).
\bibitem{LAH-12}
T. H. E. Lahtinen, K. J. A. Franke, and S. van Dijken,
Sci. Rep. {\bf 2}, 258 (2012).
\bibitem{SCH-12}
A. J. Schellekens, A. van den Brink, J. H. Franken, H. J. M. Swagten, and B. Koopmans,
Nat. Commun. {\bf 3}, 847 (2012). 
\bibitem{CHI-12}
D. Chiba, M. Kawaguchi, S. Fukami, N. Ishiwata, K. Shimamura, K. Kobayashi, and T. Ono,
Nat. Commun. {\bf 3}, 888 (2012). 
\bibitem{BAU-13}
U. Bauer, S. Emori, and G. S. D. Beach,
Nat. Nanotech. {\bf 8}, 411 (2013). 
\bibitem{LEI-13} 
N. Lei, T. Devolder, G. Agnus, P. Aubert, L. Daniel, J-V Kim, W. Zhao, T. Trypiniotis, R. P. Cowburn, C.
Chappert, D. Ravelosona, and P. Lecoeur,
Nat. Commun. {\bf 4}, 1378 (2013).
\bibitem{SCH-74} 
N. L. Schryer and L. R. Walker,
J. Appl. Phys. {\bf 45}, 5406 (1974).
\bibitem{NAK-03} 
Y. Nakatani, A. Thiaville, and J. Miltat,
Nat. Mater. {\bf 2}, 521 (2003).
\bibitem{BEA-05} 
G. S. D. Beach, C. Nistor, C. Knutson, M. Tsoi, and J. L. Erskine,
Nat. Mater. {\bf 4}, 741 (2005).
\bibitem{HAY-07} 
M. Hayashi, L. Thomas, C. Rettner, R. Moriya, and S. S. P. Parkin,
Nat. Phys. {\bf 3}, 21 (2007).
\bibitem{LAH-11-1}
T. H. E. Lahtinen, J. O. Tuomi, and S. van Dijken, 
Adv. Mater. {\bf 23}, 3187 (2011).
\bibitem{LAH-11-2}
T. H. E. Lahtinen, J. O. Tuomi, and S. van Dijken, 
IEEE Trans. Magn. {\bf 47}, 3768 (2011).
\bibitem{CHO-12}
R. V. Chopdekar, V. K. Malik, A. Fraile Rodr{\'{\i}}guez, L. Le Guyader, Y. Takamura, A. Scholl, D.
Stender, C. W. Schneider, C. Bernhard, F. Nolting, and L. J. Heyderman,
Phys. Rev. B {\bf 86}, 014408 (2012).
\bibitem{LAH-12-2}
T. H. E. Lahtinen, Y. Shirahata, L. Yao, K. J. A. Franke, G. Venkataiah, T. Taniyama, and S. van
Dijken, 
Appl. Phys. Lett. {\bf 101}, 262405 (2012).
\bibitem{HAN-00} 
R. C. O'Handley, {\it Modern Magnetic Materials: Principles and Applications} (John Wiley \& Sons, Inc.,
New York, 2000).
\bibitem{LIT-55} 
E. A. Little,
Phys. Rev. {\bf 98}, 978 (1955).
\bibitem{JO-09} 
J. Y. Jo, S. M. Yang, T. H. Kim, H. N. Lee, J.-G. Yoon, S. Park, Y. Jo, M. H. Jung, and T. W. Noh,
Phys. Rev. Lett. {\bf 102} 045701 (2009).
\bibitem{SON-13} 
J. Y. Son and S. M. Yoon,
Ceram. Int. {\bf 39}, 4031 (2013).
\bibitem{SHI-07}
Y.-H. Shin, I. Grinberg, I-W. Chen, and A. M. Rappe,
Nature {\bf 449}, 881 (2007).
\bibitem{VAN-11}
A. Vansteenkiste and B. Van de Wiele, 
J. Magn. Magn. Mater. {\bf 323}, 2585 (2011).
\bibitem{FRA-12}
K. J. A. Franke, T. H. E. Lahtinen, and S. van Dijken,
Phys. Rev. B {\bf 85}, 094423 (2012).
\bibitem{comment1} 
The lagging behind distance in the quasistatic regime $v<v_{th}$ is determined by quadratic
interpolation of the discretization points defining the center of the domain wall.
\bibitem{ZIN-11}
C. Zinoni, A. Vanhaverbeke, P. Eib, G. Salis, and R. Allenspach, Phys. Rev. Lett. {\bf 107},
207204 (2011).
\bibitem{YAN-11} 
M. Yan, C. Andreas, A. K\'akay, F. Garcia-S\'anchez, and R. Hertel,
Appl. Phys. Lett. {\bf 99}, 122505 (2011).
\end{thebibliography}
\end{document}